\documentstyle[aps,prl]{revtex} 

\begin{document}
\author{Lutfi Ozyuzer$^{1,2,}$\footnote [1]{Tel.: (630) 252 8457; 
Fax: (630) 252 7777; e-mail: ozyuzer@anl.gov}, Zikri Yusof$^{1,3}$, John F.  
Zasadzinski$^{1,3}$,\\
Ting-Wei Li$^{1}$, Dave G.  Hinks$^{1}$, and K. E.  Gray$^{1}$} 
\title{Tunneling Spectroscopy of Tl$_{2}$Ba$_{2}$CuO$_{6}$}

\address{$^{1}$ Science and Technology Center for Superconductivity 
and Materials Science Division,\\ Argonne National Laboratory, Argonne 
IL 60439} 
\address{$^{2}$ Department of Physics, Izmir Institute of 
Technology, TR-35230 Izmir, Turkey}
\address{$^{3}$ Science and Technology Center for Superconductivity,\\
	Illinois Institute of Technology, Chicago, IL 60616}
\maketitle

\begin{abstract}

New results from tunneling spectroscopies on near optimally-doped single 
crystals of Tl$_{2}$Ba$_{2}$CuO$_{6}$ (Tl-2201) junctions are 
presented.  The superconductor-insulator-normal metal (SIN) tunnel 
junctions are obtained using the point-contact technique with a Au 
tip.  The tunneling conductances reproducibly show a sharp cusp-like 
subgap, prominent quasiparticle peaks with a consistent asymmetry, and weakly 
decreasing backgrounds.  A rigorous analysis of the SIN tunneling data 
is performed using two different models for the $d_{x^{2}-y^{2}}$ 
($d$-wave) density of states (DOS).  Based on these and earlier 
results, the tunneling DOS of Tl-2201 have exhibited the most 
reproducible data that are consistent with a $d$-wave gap symmetry.  
We show that the dip feature at $2\Delta$ that is clearly seen in SIN 
tunneling data of Bi$_{2}$Sr$_{2}$CaCu$_{2}$O$_{8+\delta}$ is also 
present in Tl-2201, but at a weaker level.  The gap values for 
crystals with a bulk $T_{c} = 86$ K are in the range of 19-25 meV.


\end{abstract}
\section{INTRODUCTION}

Tunneling spectroscopy has revealed the complex characteristics 
of high-$T_{c}$ superconductors (HTS's). Tunneling spectra on
Bi$_{2}$Sr$_{2}$CaCu$_{2}$O$_{8+\delta}$ (Bi-2212)\cite{yannick,renner}, 
Bi$_{2}$Sr$_{2}$CuO$_{x}$ \cite{paola} and 
HgBa$_{2}$CuO$_{4}$ (Hg-1201)\cite{chen,wei}, have shown both symmetric 
and asymmetric tunneling conductance peaks, and variable subgap features that 
range from sharp cusp-like to flat, BCS-like.  Additionally, tunneling 
experiments on YBa$_{2}$Cu$_{3}$O$_{7}$ (YBCO) and Bi-2212 in certain 
crystal orientations have also shown the presence of zero-bias peaks 
in the conductance data.\cite{covington,wei2,sinha} In Bi-2212 there 
exists a prominent dip feature (at eV$\sim 2\Delta$) that is 
asymmetric with bias voltage, being much stronger for a polarity that 
corresponds to removal of quasiparticles from the superconductor.  These 
unusual observations have made it difficult to 
properly analyze the results of tunneling experiments and have 
complicated the deduction of important properties of HTS's such as the 
pairing symmetry.

There is an emerging consensus that the predominant pairing symmetry 
in hole-doped HTS cuprates is $d_{x^{2}-y^{2}}$ ($d$-wave).  Evidence from 
tricrystal ring\cite{kirtley} experiments points to pure $d$-wave for 
YBCO.  Grain boundary \cite{kouznetsov} and scanning 
tunneling microscopy (STM) \cite{wei2} 
junctions indicate a small $s$-wave contribution to the $d$-wave 
symmetry on YBCO which was attributed to the orthorhombicity of YBCO.  
Tunneling \cite{huangNature} and penetration depth \cite{Wu} measurements of 
electron-doped Nd$_{2-x}$Ce$_{x}$CuO$_{4}$ are compatible with s-wave 
symmetry.  Another well-studied, hole doped, HTS is Bi-2212 because of 
the availability of high quality single crystals, and the ability to 
easily cleave this crystal along the {\it a-b} plane.  Results from 
angle resolved photoemission spectroscopy (ARPES) indicate an 
anisotropic gap with a minimum in the ($\pi$,$\pi$) direction that is 
consistent with $d$-wave symmetry\cite{shen}.  Furthermore, results 
from ARPES and STM have exhibited spectral features that are also 
observed in PCT, such as quasiparticle peak, dip and hump.  They have 
also shown the puzzling feature of an increasing energy gap size 
with decreasing doping concentration in Bi-2212.\cite{yannick,renner} 
DeWilde et al.  \cite{yannick} have shown that the study on Bi-2212 
using three different techniques (PCT, break-junction, and STM) 
can produce very similar results as far as gap size, dip structure, and 
subgap shape are concerned, even when the resistance of the STM 
junction was of the order of G$\Omega$ while it was the order of 
1k$\Omega$-100k$\Omega$ for PCT and break junction.  However, point contact 
tunneling (PCT) results also occasionally show a flat subgap 
structure\cite{yannick} which is not easily reconciled with $d$-wave 
symmetry.

Quasiparticle tunneling has failed to definitely reveal the pairing
symmetry in Hg-1201. PCT on polycrystal samples of this 
HTS seems to show a density of states (DOS) that is flat near zero-bias, 
consistent with $s$-wave symmetry\cite{chen,jeong}, whereas Wei {\it 
et al}.  \cite{wei} claim that STM measurements on the same HTS seems 
to be consistent with a $d$-wave gap symmetry.  While it has been 
shown that tunneling directionality effects can produce a flat subgap 
conductance with a $d$-wave gap,\cite{yusof} there is no obvious 
physical mechanism for preferred tunneling directions.  It is 
therefore more likely that the sporadic observations of flat subgap 
conductances in HTS simply adds fuel to the debate over pairing 
symmetry.

Experimental evidence of $d$-wave pairing symmetry on Tl-2201 is more 
convincing.  Results from tricrystal ring experiments indicate a pure
$d$-wave pairing\cite{tsuei}, although admixture of $d$ and $s$-wave 
pairing is also interpreted from in-plane torque anisotropy 
experiments.\cite{willemin} We have earlier reported \cite{lutfi} the 
tunneling studies of optimally-doped Tl$_{2}$Ba$_{2}$CuO$_{6}$ crystals
(Tl-2201) with $T_{c}=91$ K which clearly and reproducibly showed 
a tunneling DOS that is consistent with a momentum-averaged 
$d$-wave gap symmetry. In that report (Ref. \cite{lutfi}), 
our analysis of the superconductor-insulator-normal metal 
(SIN) tunneling conductance was
somewhat primitive, utilizing a simple model for the $d$-wave DOS.
In this report, we present additional  tunneling data on Tl-2201 
crystals with $T_{c} = 86$ K that have been synthesized using a 
different technique than the one described in Ref. \cite{lutfi}. 
The location of the quasiparticle peaks in the SIN conductance
data are consistent with the ones in our earlier report and all 
of the data again display the cusp feature at zero bias.  However, 
here we present a more exhaustive treatment of many junctions, with a 
wide range of junction conductance ($\sim$0.1 mS-2 mS).  We have also 
performed a more rigorous analysis of the SIN conductance data using 
two different models for the tunneling DOS with a $d$-wave gap.  We 
again find good agreement with $d$-wave symmetry.

Some of the SIN data display a weak dip feature at eV$\sim 2\Delta$. We have 
generated superconductor-insulator-superconductor (SIS) conductance 
curves using the SIN data.  The resulting SIS curves display the 
characteristic dip features at nearly $3\Delta$ that are consistent 
with those observed in the SIS tunneling conductance of Bi-2212.  This 
is the first study to clearly indicate that the dip feature is present 
in the SIN conductance data of Tl-2201, but with a smaller magnitude 
than observed in Bi-2212.  Further comparison with tunneling data of 
Bi-2212 reveals that while the bulk $T_{c}$ of Bi-2212 and Tl-2201 are 
approximately the same, the magnitude of the typical energy gap of 
Tl-2201 is smaller.  The origin of this discrepancy is still unknown 
at present, but some insight has been gained with this study.  We note first 
that the largest gaps found for the Tl-2201 ($\Delta$=25 meV) are 
close to those of Bi-2212 when both have the same T$_{c}$=86 K.  
Furthermore, due to the strong dependence of the gap magnitude on 
doping concentration \cite{yannick} in Bi-2212, we suggest that the 
smaller gap values in Tl-2201 may be due to a surface that is slightly 
overdoped.

\section{EXPERIMENTAL PROCEDURE AND RESULTS}

Tl-2201 has a tetragonal crystal structure with a single $Cu-O$ layer
per unit cell which is relatively simple when compared to the bilayer
and trilayer high-$T_{c}$ superconductors.  However, 
Tl$_{2}$Ba$_{2}$Ca$_{2-n}$Cu$_{2}$O$_{2n+3}$ (n=1, 2, and 3) family
is very sensitive to thallium and oxygen content which influences the 
structure and superconducting properties.\cite{Storm} The 
optimally-doped compound of Tl-2201 has a $T_{c}$ of approximately 91 
K and this value can be reduced to zero on the overdoped side by oxygen
annealing.\cite{shimakawa}

The Tl-2201 single crystals were grown from a flux in an alumina 
crucible with an alumina lid, sealed to avoid loss of thallium oxide. 
Tl$_{2}$O$_{3}$, BaO$_{2}$ and CuO powders were mixed at the atomic 
ratio of Tl:Ba:Cu=2.2:2:2 using excess Tl$_{2}$O$_{3}$ and CuO as the 
flux. The crucibles, containing about 50 g of charge, were loaded 
in a vertical tube furnace and heated rapidly to 925-950$^{o}$C. 
This temperature was held for 1/2 hour. The furnace was then cooled 
at 5 $^{o}$C/h to 875$^{o}$C, and finally cooled to room 
temperature. The crystals were platelet-shaped, with a basal plane 
area of about 1 mm$^{2}$ and a thickness along the $c$-axis varying 
between 20-100$\mu$m.  The critical temperature of the samples is 
determined by {\it ac} magnetization measurements.

The experimental setup of our PCT system is designed for data 
collection over a large range of sample temperature.  In addition to 
this feature, tunneling measurements can also be performed in high 
magnetic fields, up to 6 T.  The details of the measurement system can 
be found elsewhere.\cite{system} Cleaved single crystal samples of 
Tl-2201 usually have shiny surfaces in the {\it a-b} plane.  Each is 
mounted on a substrate using an epoxy so that the tip approaches 
nominally along the $c$-axis.  The electrical leads are connected to 
two sides of the sample by using silver paint.  Non-superconducting 
{\it Au} is used as a counter-electrode.  It is mechanically and 
chemically cleaned before each run.

While the differential micrometer driven tip approaches the sample, 
the $I(V)$ signal is continuously monitored on an oscilloscope until an
acceptable tunnel junction is obtained, i.e. one which displays an
obvious superconducting gap feature. All tunnel junctions are
initially formed at 4.2 K to prevent any  sample surface deterioration.
First derivative measurements, $\sigma = dI/dV$, were obtained using 
a Kelvin bridge circuit with the usual lock-in procedure. $I(V)$ 
and $dI/dV$ are simultaneously plotted on a chart and digitally 
recorded on a computer.  DeWilde {\it et al.}\cite{yannick} have 
shown that tunneling results on Bi-2212 using PCT can produce 
results that are consistent with those obtained using STM.

In contrast to other surface sensitive experimental methods such as
STM, ARPES, Raman, and auger techniques, the advantage of the point 
contact method for cuprates is that the tip can be used to scrape, 
clean, and in some cases cleave the surface.  The tip often can 
penetrate through the surface and reach the bulk of the crystal.  
The cleaving of the surface sometimes results in the formation of SIS 
junctions, as in the case of Bi-2212.  This happens when a piece of the 
HTS crystal attaches itself to the tip, forming an ohmic contact.  As 
the tip is retracted, the piece forms an SIS break junction with the bulk 
crystal\cite{lutfi2}.  Unlike Bi-2212, Tl-2201 has stronger bonds 
between planes and consequently SIS junctions could not be formed this 
way.

Figure 1 shows the conductances of eight junctions on three
different Tl-2201 crystals, each with a bulk $T_{c}$ near 86 K. These
junctions are representative of a larger set of data and they
demonstrate several characteristics that are typical for PCT tunneling
in Tl-2201. Each junction exhibits a single energy gap feature with
conductance peaks at $|V|= 20-25$ mV.  The voltage is that of the 
sample respect to the tip and thus negative bias corresponds to 
removal of electrons from the superconductor.  There is a characteristic 
asymmetry in the conductance peaks such that the negative bias peak is 
higher than the one at positive bias.  This type of asymmetry has also 
been seen in PCT and STM studies of Bi-2212, most consistently in 
overdoped samples.  It has been pointed out that this conductance peak 
asymmetry may be a signature of the $d$-wave pairing.\cite{yusof}

The background conductances for $|eV|>\Delta$ are generally weakly
decreasing with bias similar to that in Bi-2212, and it is these type of
junctions that exhibit the largest peak height to background (PHB)
ratio. A few junctions show a flat or slightly increasing background
with a smaller PHB ratio. This implies that the decreasing background is
an intrinsic property of the quasiparticle DOS. While such a feature may
be due to the underlying band structure DOS, we note the absence of any
van Hove singularity (VHS) in these data as well as earlier PCT data on
Tl-2201. All of the junctions exhibit a cusp-like feature at zero bias
which is characteristic of a $d$-wave DOS.

\section{THEORETICAL MODEL}

The tunneling data are analyzed with two different methods. For 
Model I, the superconducting data are first normalized by 
constructing a "normal state" conductance obtained by 
fitting the high bias data to a third order polynomial.  The normalized 
conductance data are compared to a weighted momentum averaged $d$-wave 
DOS,
\[
N(E)=\int f(\theta )\frac{E-i\Gamma}{\sqrt{(E-i\Gamma)^{2}-\Delta (\theta 
)^{2}}}d\theta.
\]
Here $\Gamma$ is a lifetime broadening factor,
$f(\theta )$, \cite{manabe} is an angular
weighting function, and $\Delta(\theta )=\Delta _{o}\cos(2\theta )$
represents the $d$-wave gap symmetry expected from a mean-field 
BCS-type interaction.\cite{maki} This model is used because it allows
for a quick estimation of the gap value and as we will show, gives 
an excellent fit to the data. The inclusion of the weighting 
function allows for a better fit with the experimental data 
in the gap region than with the non-weighted average as was done 
previously.\cite{lutfi} Here, a weighting function $f(\theta)=1+0.4 
cos(4\theta)$ was used which imposes a preferential angular selection of the 
DOS along the absolute maximum of the $d$-wave gap and tapers off 
towards the nodes of the gap.  This is a rather weak directional 
function since the minimum of $f(\theta)$ along the nodes of the 
$d$-wave gap is still non-negligible.

The second method (Model II) makes no attempt to normalize out any 
background conductance.  Rather, an attempt is made to fit the entire 
spectrum by including a band structure, tunneling 
matrix element, and $d$-wave gap symmetry.\cite{yusof}  The tunneling DOS
is calculated using the single particle Green's function,

\begin{equation}
N(E)=-\frac{1}{\pi }%
\mathop{\rm Im}%
\sum_{{\bf \ k}}|T_{{\bf k}}|^{2}G({\bf k},E)
\end{equation}

For the superconducting state,

\[
G({\bf k},E)=\frac{u_{k}^{2}}{E-E_{k}+i\Gamma }+\frac{v_{k}^{2}}{%
E+E_{k}+i\Gamma } 
\]
where $u_{k}^{2}$ and $v_{k}^{2}$ are the usual coherence factors, $\Gamma
$
is the quasiparticle lifetime broadening factor, and $E_{k}=\sqrt{|\Delta
(%
{\bf k})|^{2}+\xi _{{\bf k}}^{2}}$ with the gap function for d-wave
symmetry 
$\Delta ({\bf k})=\Delta _{o}[\cos (k_{x}a)-\cos (k_{y}a)]/2$ . The
tunneling matrix element $\left| T_{{\bf k}}\right| ^{2}$ is written as

\[
\left| T_{{\bf k}}\right| ^{2}=v_{g}D({\bf k}) 
\]
where $v_{g}$ is the group velocity defined as $v_{g}=\left| \nabla _{{\bf
k}%
}\xi _{{\bf k}}\cdot {\bf n}\right| $ and {\it D}({\bf k}) is the
directionality function that has the form\cite{klemm}

\begin{equation}
D({\bf k})=\exp \left[ -\frac{k^{2}-({\bf k}\cdot {\bf n})^{2}}{({\bf k}%
\cdot {\bf n})^{2}\theta _{o}^{2}}\right] 
\end{equation}
Here the unit vector {\bf n} defines the tunneling direction, which is
perpendicular to the plane of the junction, whereas $\theta _{o}$
corresponds to the angular spread in {\it k}-space of the quasiparticle
momenta with respect to {\bf n} that has a non-negligible tunneling
probability.

The band structure for the $Cu-O$ plane 
extracted from ARPES measurements on Bi-2212 is used.\cite{norman} 
Presumably, other than the exact value of the chemical potential, the 
band structure for Tl-2201\cite{singh} should have the same generic 
features as the extracted band structure from Bi-2212.  Unlike a 
similar analysis done in our earlier report,\cite{lutfi} the presence of the 
VHS is not artificially removed by using a large chemical potential.  
Rather, the presence of the VHS is effectively diminished by the group 
velocity factor from the tunneling matrix 
element.\cite{yusof,harrison} Here, the value of the chemical 
potential has been altered slightly to produce the best comparison to 
the experimental data.  The tunneling DOS from this model is compared 
directly to the experimental conductance data using a constant scaling 
factor.  An interesting aspect of this second model is the robust 
asymmetric quasiparticle peaks in the tunneling DOS.  This asymmetry, 
which has the higher peak in the filled states, is a direct 
consequence of the $d$-wave gap symmetry and directionality in the 
model.  As will be shown, this result is consistent with our 
experimental tunneling data.

\section{ANALYSIS AND DISCUSSION}

Figures 2 and 3 present two representative SIN tunneling conductances 
of Tl-2201.  Figure 2 shows Junction E of Fig.\ 1, while Figure 3 is
an additional conductance curve (Junction J) not shown in Fig.\ 1 that has 
a very high peak height to background ratio.  As illustrated in Figs.\ 
2(a) and 3(a), the SIN conductance data consistently display the 
sharp, cusp-like subgap feature, weakly decreasing background, and 
conductance peaks that are either weakly or strongly asymmetric, with 
the higher peak on the negative bias side.  The presence of these 
features and the overall shape of the conductance data are very 
similar to the conductance data of Fig.\ 1 in our earlier 
report.\cite{lutfi}

To compare the two data sets to Model I, the SIN conductance data 
are first normalized by dividing through with an extrapolated 
normal state conductance curve which is shown as the solid 
line in Fig. 2(a) and 3(a).\cite{lutfi} The normalized conductances 
are then compared to the DOS obtained from Model I as shown in Fig. 
2(b) and 3(b). Other than a remaining conductance peak asymmetry 
and somewhat broader experimental peaks, the model DOS shows a 
remarkably good overall fit in the gap region with the experimental 
DOS. Notice that while the process of normalization has reduced 
the degree of asymmetry of the conductance peaks in both data 
sets, it has not eliminated it.  This proves that the asymmetry is not 
a consequence of the background.

The comparison of Model II with the unnormalized tunneling data is
shown in Fig. 2(c) and 3(c). The most striking observation is the 
model's ability to reproduce the peak asymmetry that is seen in 
the data. As was shown in Ref. \cite{yusof}, this type of 
asymmetry with the higher peak in the filled states is a 
robust property of $d$-wave gap symmetry and directional 
tunneling. The strength of the directionality here is
defined by $\theta_{o}$ and the values
used to compare both experimental data here are
considerably larger than the ones used to analyze the Bi-2212
data.\cite{yusof} This implies that these two data sets are best fit 
with a weak directional tunneling processes which is consistent with 
the type of weighting function $f(\theta)$ in Model I.

As in Bi-2212, Model II could not accurately reproduce the background 
conductance although it does show the generic decreasing background 
seen in both data sets.  This may also be due to the fact that we are 
not using the exact normal state band structure for Tl-2201 in the 
model.  Model II also produces a poorer agreement with the subgap data 
which might be due to the particular choice of directionality function 
used.  Note that the values of the energy gap from both models are 
very close to each other, with the Model I having a slightly lower gap 
values than Model II.

We would like to point out that attempts at comparing the 
normalized experimental data with just a pure $d$-wave 
gap symmetry without the $f(\theta)$ weighting function 
in the Model I led to a poorer fit to the data.  Considering this, we 
reanalyze some of the SIN tunneling data from our previous report 
(Junction B and C in Fig.  4 of Ref.  \cite{lutfi} and relabeled here 
as Junction B' and C' respectively).  We have restricted this analysis 
by using only Model I.  Figure 4(a) shows the raw SIN conductance data 
of Junction B' and the estimated normal state conductance used to 
obtain the normalized data.  This normalized curve is shown in Fig.  
4(b) along with the comparison to Model I.  The model produces a 
better fit in the gap region (with identical gap value) when compared 
to our earlier fit.  This procedure is repeated for Junction C' as 
shown in Fig.  4(c) (which has been normalized by a constant).  In 
this case, the overall fit is only slightly improved over the one we 
reported earlier, with an identical gap value.

One of the distinct features of the tunneling DOS in Bi-2212 is the 
strong dip beyond the quasiparticle peak\cite{SPIE} in the occupied 
states.  This feature is clearly seen in SIN conductance data of 
Bi-2212 from both STM and PCT.\cite{yannick} Furthermore, this dip 
feature is enhanced in the superconductor-insulator-superconductor 
(SIS) junction.  This is apparent from the break junction tunneling 
data in Ref.  \cite{yannick}.  Our SIN conductance data of Tl-2201 
from this work and in our previous report do not seem to distinctively 
show the same dip features, although there is evidence of a weak dip 
feature in the normalized data of Figs.  2 and 4 as well as junctions 
C and G in Fig. 1.  We explore this issue further by generating SIS conductance 
curves from the raw data of Junction E and Junction B' which should 
enhance any dip feature that may exists in these SIN data.  As shown 
in Fig.  5, both conductance data generate SIS curves that are 
qualitatively similar to the experimental SIS tunneling data of 
Bi-2212.\cite{yannick}  Both curves clearly display the prominent dip 
features located at slightly less than 3$\Delta$.  This indicates that 
the dip feature is also present in the SIN conductance data of Tl-2201 
but at a smaller amplitude than the ones observed in the SIN data of 
Bi-2212.

Another significant difference between the tunneling data of Bi-2212 
and Tl-2201 is the magnitude of the superconducting gap and 
this might be related to the dip feature discussed above. 
The optimally-doped Bi-2212 which has a $T_{c}$ of 93-95 K has an 
energy gap in the range of 37-38 meV.\cite{yannick} Due to the high 
reproducibility of the gap value for Bi-2212, it is presumed that the 
gap value for Bi-2212 is consistent with its $T_{c}$.   
Tl-2201 which has a bulk $T_{c}$ of 86 K in this 
study and 91 K in the previous study, has an energy gap in the range 
of 19-25 meV.  This value is considerably less than the energy gap of 
Bi-2212 even though bulk $T_{c}$ for both cuprates are roughly 
the same.  This discrepancy raises an important question in HTS cuprates, 
namely the relationship between gap size $\Delta(T=0)$ and $T_{c}$.
The unusual $\Delta$ versus doping in Bi-2212, which violates mean-field 
theory, strongly suggest that $T_{c}$ is a phase coherence 
temperature.  In this picture, there are strong superconducting 
fluctuations above $T_{c}$ and presumably the ability of each HTS to 
support such fluctuations depends on structural parameters, anistropy 
and the degree of 2-dimensionality.  It is thus possible that there is 
no universal relationship between $\Delta$ and $T_{c}$ for all HTS.
If there exists a universal relationship between 
these two parameters for HTS cuprates as is approximately the case for 
conventional superconductors, then the difference in the gap size 
between these two HTS's needs another explanation.  It is possible, 
due to the strength of the interplane bonding, that the tunneling 
measurement is probing predominantly the surface of the Tl-2201 
crystals which has been exposed to air and may have properties 
different from the bulk.  This raises the possibility that the surface of 
Tl-2201 may be slightly overdoped, which results in a smaller gap 
size.  When Bi-2212 is annealed in air it has a $T_{c} \sim$86 K and is 
slightly overdoped.  This is the equilibrium oxygen doping level at 
atmospheric conditions and a similar situation is found for Tl-2201.  
Air annealed Tl-2201 has a $T_{c} \sim$82 K.  Therefore, air-exposed 
Tl-2201 will have a tendency for the surface to be somewhat overdoped 
by coming to equilibrium with atmospheric conditions.  We are then 
suggesting that when the sample is cooled down to 4.2 K, there are no 
changes in the surface concentration.  Of course we have no proof of 
this.  If there are changes in the surface concentrations upon cooling 
in vacuum, then these changes are highly reproducible because both 
Bi-2212 and Tl-2201 display highly reproducible spectra and gap 
values.

Furthermore, the strength of 
the dip feature seems to indicate that the surface is slightly 
overdoped.  In Bi-2212, tunneling conductances exhibiting gap sizes of 
35-40 meV exhibit dip strength of approximately 80\% of the
background
conductance.  For smaller gaps in the range of 15-20 meV (which are from 
overdoped Bi-2212), the dip strength is approximately 
10\%.\cite{yannick}
This is consistent with what is observed in Fig.  5 for Tl-2201 and 
seems to support our argument that the surface of Tl-2201 crystals we 
measured is slightly overdoped. This however, is still speculation and 
requires further detailed study to account for the apparent gap size 
discrepancy.  We note that preliminary temperature dependent data 
indicate that junctions which exhibit small gaps ($\sim$ 20 meV) also 
show a strong smearing out of the gap feature at a temperature below 
the bulk T$_{c}$.

To summarize, we have performed SIN tunneling junction measurements on 
single crystals of Tl-2201 with bulk $T_{c}$ of 86 K.  The 
conductance data obtained reproducibly show cusp-like subgap features, 
asymmetric conductance peaks and weakly decreasing backgrounds.  These 
observations are consistent with our earlier report on Tl-2201 with a 
$T_{c}$ of 91 K that were synthesized in a different manner.  The 
present data are fit reasonably well with two different models using the 
$d$-wave gap symmetry.  The need for a weighting function in Model I 
and the prominent asymmetry of the data which is reproduced in Model 
II seem to indicate that the tunneling process in these cases may have 
a weak preferential tunneling direction centered at or near the 
absolute maximum of the $d$-wave gap.  The magnitude of the 
superconducting gap for this cuprate is noticeably smaller than the 
gap size of optimally-doped Bi-2212 that has similar $T_{c}$.  The 
existence of a universal relationship between the superconducting gap 
size and $T_{c}$ is still undetermined, and therefore the origin of 
the discrepancy between the gap size of these two cuprates is still 
uncertain.

\section{ACKNOWLEDGMENTS}

This work was partially supported by U.S. Department of 
Energy, Division of Basic Energy Sciences-Material Sciences 
under contract No. W-31-109-ENG-38, and the National Science 
Foundation, Office of Science and Technology  Centers 
under contract No.  DMR 91-20000.  Z.Y. acknowledges
support from the Division of Educational Programs, 
Argonne National Laboratory.

\newpage

Fig. 1. Tunneling conductances of eight junctions on three different 
Tl-2201 crystals, each with a bulk $T_{c}$ near 86 K.  Junction A, B, 
C, E, and F have been shifted vertically by 1.5, 1.2, 0.7, 0.3, and 0.1 mS 
respectively for clarity in their own scales.

Fig. 2. (a) SIN tunneling conductance of Junction \ E (circles) at 4.2 K
and the estimated normal state conductance (line). (b) Comparison of the
normalized SIN conductance with Model I. The inset shows the angular
weighting function $f(\theta)$. (c) Comparison of the unnormalized
conductance with Model II. Refer to Ref. \cite{yusof} for
definitions of variables. $c_{o}$, which corresponds to the chemical
potential, has been changed to 0.1585 eV for all comparisons done in
this paper.

Fig. 3. (a) SIN tunneling conductance of Junction J (circles) at 4.2 K
and the estimated normal state conductance (line). (b) Comparison of the
normalized SIN conductance with Model I. (c) Comparison of the
unnormalized conductance with Model II.

Fig. 4. (a) SIN tunneling conductance of Junction \ B' (circles) at 4.2 K
and the estimated normal state conductance (line). (b) Comparison of the
normalized SIN conductance with Model I. (c) Comparison of the
normalized SIN conductance of Junction \ C' with Model I. The tunneling
conductance has been normalized by a constant.

Fig. 5. SIS conductance curves generated from the unnormalized
SIN conductance curves of Junction \ E and \ B'. Each SIS curve shows the
prominent dip feature at nearly 3$\Delta$.

\end{document}